# AUMATED MULTI-DOMAIN ENGINEERING DESIGN THOUGH LINEAR GRAPS AND GENETIC PROGRAMMING

Eric McCormick*, Haoxiang Lang* and Clarence W. de Silva**

*GRASP Laboratory, Department of Automotive and Mechatronics, Ontario Tech University, Oshawa, ON, Canada.

Emails: eric.mccormick@ontariotechu.net, haoxiang.lang@ontariotechu.ca

**Industrial Automation Laboratory, Department of Mechanical Engineering, The University of British Columbia, Vancouver, BC, Canada.

Email: desilva@mech.ubc.ca

## Abstract

This paper proposes a methodology of integrating the Linear Graph (LG) approach with Genetic Programming (GP) for generating an automated multi-domain engineering design approach by using the in-house developed LG MATLAB toolbox and the GP toolbox in MATLAB. The necessary background for the development are presented, and the methodology used in this work to facilitate the construction and evaluation of filter circuits, using LG models, is described. Designing electronic filter circuits through an evolution from electronic components to the completed circuits is demonstrated. The topology and component values of three types of filter circuits: low pass, high pass, and band pass, are designed through this evolutionary approach, for various cut-off frequencies. Furthermore, the paper demonstrates through examples of these evolved filter circuits, the combined GP and LG approach is successful in constructing high order filter circuits that are capable of attenuating undesired frequencies while maintaining desirable ones. The work presented in the paper is a key step towards the integration of LG modeling, through the use of the LGtheory MATLAB Toolbox, with machine learning techniques for the automated design of dynamic multi-domain mechatronic systems.

# 1. Introduction

From voice and facial recognition software to self driving vehicles and beyond, the applications of artificial intelligence (AI) and machine learning techniques to various aspects of our lives is an ever-growing field of research. One such up and coming focus of this research field is the use of machine learning in the design of mechatronic engineering systems. Similar to AI and machine learning, the field of mechatronic engineering, involving the design of multi-energy-domain systems, is one that is becoming ever more prevalent. Systems that were once purely single-domain in nature (mechanical, electrical, hydraulic, thermal, etc.) are now being improved immensely through their integration, by the combination of energy-domain sub-systems, with the benefits of more robust functionality, improved efficiency and reliability, enhanced ability for system optimization, and accurate control.

The Linear Graph (LG) Modeling approach, performed in the present work through the use of the LGtheory MATLAB Toolbox developed in the GRASP lab at Ontario Tech University[1], is a method of dynamic system modeling which is derived from Graph Theory. The origin of LG may be traced back to the 18$^{th}$ century, with the work of Leonhard Euler in order to solve the topological problem of the Seven Bridges of Königsberg [2]. The first formal formulation of the LG Theory and its application in the modeling of engineering systems was during the 1950s by Henry M. Paynter (1961) at MIT, particularly for the purposes of evaluating large electrical networks [3]. At that time, methods of evaluating systems of various energy domains utilized different methods depending on the domain in question; the work of Koenig and Blackwell (1960) presented in [4] was a key steps towards expanding the use of LG Theory to various energy domains beyond electrical systems. Through decades of development, LG Theory has been expanded to the four primary energy domains: electrical, mechanical translation and rotational, hydraulic/fluid, and thermal [5], and also various additional application areas such as multi-body systems [2,6].

The genetic programming (GP) approach, which is implemented in the present work through the use of the GPLAB MATLAB Toolbox [7], is a machine learning and automatic programming technique which genetically evolves computer programs in the form of tree structures as a method of problem solving. Although the idea of evolving computer programs had been theorized for many years before, the paradigms that govern the GP process were formally introduced by John Koza in 1990 [8]. Since its introduction, numerous applications of GP have been explored in various fields such as engineering, computer science, biomedical, and chemistry [9]. Applications of GP in engineering design include pre-existing examples of filter circuit synthesis as demonstrated in [10-12], but the use of GP along with LG theory for the evolutionary design of multi-domain mechatronic systems is still rather scarce, with examples of it being rather limited [13]. Although the present paper is only concerned with the evolution of filter circuits in the electrical-domain, by demonstrating the successful implementation of the LGtheory MATLAB Toolbox for the purposes of automated design, this work may then be extended in a straightforward manner to systems encompassing multiple energy domains.

The purpose of the present paper is to demonstrate the potential of combining the two fields, machine learning and mechatronics design, through the evolutionary design methodology of engineering systems to achieve computer automated design of multi-domain engineering systems. In this paper, the general concepts of LG and GP are introduced in Section 2, along with the development of the LG-based MATLAB toolbox. Detailed examples are given for converting physical systems into linear graph models and eventually to the MATLAB code using the developed MATLAB toolbox called LGtheory. With the integration of the GP and LG approaches through the combination of LGthoery and GPLab toolboxes, an evolutionary design method is proposed with detailed explanations of the terminology and design steps. In Section 3, examples of designing filters are presented to demonstrate the application of the proposed method. Results of the implementations are provided and discussed in Section 4. Conclusion are drawn in the last section of the paper.

## 2. Background

### 2.1. Linear Graph Modeling

The LG modeling approach provides a method of representing and evaluating complex dynamic systems through the use of a simplistic graphical and topological representation in order to facilitate the derivation of state-space models. The systematic, unique, unified, and integrated nature of the LG modeling approach allows for a robust modeling method, which produces a unique solution through the application of methodologies analogously across multiple energy domains (i.e., a mechatronic system). This means that a multi-domain system is evaluated in its entirety (not sequentially, like with other methods), while applying the same basic procedures for each energy domain contained within the system to produce a unique state-space model [14].

The LG modeling approach represents dynamic systems through two main components: 1) branches: line segments used to represent the elements of the system; and 2) nodes: a point representing the physical connections between the system elements. Similarly, two main variable types are considered in the LG modeling approach: across-variables, denoted generally as $v$, which are the variables measured "across" an element (such as voltage across a resistor or velocity across an inertia element); and through-variables, denoted generally as $f$, which are the variables measured "through" an element (such as current through a resistor or force through a spring). The product of the Across- and Through-variables of a system element determines the power flow of that element.

The three primary passive single-port element types represented in the LG modeling approach include A-type, T-type, and D-type elements. A- and T-type elements are energy storage elements whose energy storage is defined by their across-variable and through-variable, respectively. Examples of these elements include capacitors and inertial elements for A-types, and inductors and springs for T-types. D-type elements are energy dissipating elements whose energy dissipation is define by either their across or through-variables. Examples for this element type include resistors and dampers. Likewise, there are two source element types, A-type and T-type source elements, which provide energy to the system in the form of their respective variable type.

There are two additional passive two-port element types, transformers and gyrators, which are non-dissipative elements that either convert the magnitude of the variables acting upon it, or convert the energy-type to that of a different energy-domain. An example of a magnitude converting element is an ideal electrical transformer, where the magnitude of the input current is traded off for the magnitude of the output voltage, or vise versa. An example of a domain converting element is a DC motor, where electrical energy input into the element is converted to mechanical energy at the output. In the case of a transformer element, the input variable type is related to the same output variable type, while in the case of a gyrator element, the input variable type is related to the opposite variable type.

The constitutive equations of each aforementioned element type is presented in Table 1.

Table 1: Constitutive equations of single- and two-port elements

| Element | Constitutive Equations |
| --- | --- |

| | | |
|---|---|---|
| Generalized A-type | $f = C_i \dfrac{dv}{dt}$ | |
| Generalized T-type | $v = L_i \dfrac{df}{dt}$ | |
| Generalized D-type | $f = \dfrac{1}{R_i} v$ | $v = R_i f$ |
| Transformer | $v_1 = TF v_2$ | $f_1 = -\left(\dfrac{1}{TF}\right) f_2$ |
| Gyrator | $v_1 = GY f_2$ | $f_1 = -\left(\dfrac{1}{GY}\right) v_2$ |

For more information of LG modeling, the references [5][14][15] provide a fundamental formulation of the LG modeling approach, introduce the basic concepts of LG theory, and several examples of various single and multi-domain systems.

### 2.1.1. LG Modeling Example

The following example of a hydraulically actuated mechanical system from [5] will be evaluated using the LGtheory MATLAB Toolbox developed in our lab [1]. This system consists of a pump, which transmits fluid through a resistive pipe into a hydraulic piston. This piston then converts the hydraulic pressure into force in the mechanical translational domain in order to actuate a mass attached to a spring, which slides on a resistive surface. The schematic diagram and the LG model of this system are shown in Fig. 1.

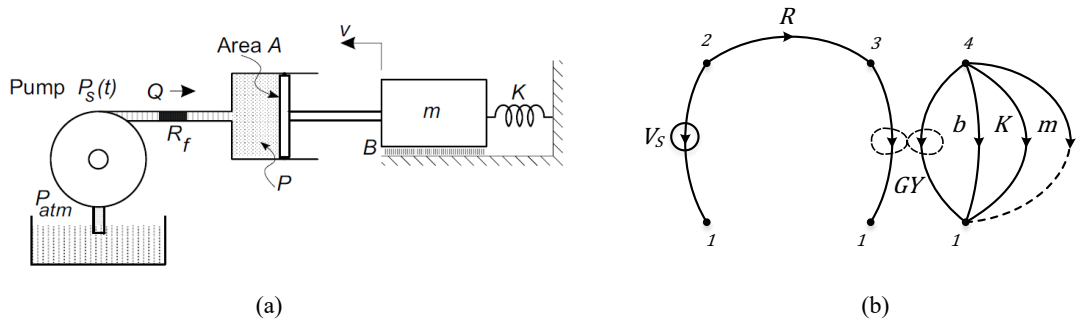

(a)           (b)

**Fig. 1** (a) Schematic model [5] and (b) LG model of the hydraulically actuated mechanical system

The required inputs to the LGtheory MATLAB Toolbox for this system are as follows:

```
S = [2 2 3 4 4 4 4];
T = [1 3 1 1 1 1 1];
Type = [1 5 4 4 5 6 2];
Domain = [4 4 4 2 2 2];
syms s R A_p b K m
Var_Names = [s R 1/A_p 1/A_p b K m];
syms v_m(t)
y = [v_m(t)];
[A,B,C,D,F,x,u] = LGtheory(S,T,Type,Domain,Var_Names,y);
```

An updated input method is now utilized by the LGtheory Toolbox since its introduction [1]. The user is no longer required to input the incidence matrix form of the LG model; instead, the user defines the model topology through the Source (**S**) and Target (**T**) vectors. The column-wise index values of the Source vector inform the program which nodes the corresponding elements are leaving, while the index values of the Target vector inform the program which nodes the elements are entering.

The Type and Domain vectors are thus formed using Table 2 with the corresponding index values related to each element.

The Var_Names vector allows the user to define the variable names for each element. These names can either be individual symbolic variables or expressions containing symbolic variables, such as the gyrator ratio of the piston being equal to the reciprocal of the piston's cross-sectional area ($GY = 1/A_p$), as seen in the example inputs above.

The user then specifies the output variables of interest in the output vector (y); for this example, the variable of interest is the velocity of the mass element, $v_m(t)$.

Table 2: Index values for element types and energy domains.

| Index | Element Type | Index | Energy Domain |
|---|---|---|---|
| 1 | A-Source | 1 | Electrical |
| 2 | A-Type Element | 2 | Mech. Translational |
| 3 | Transformer | 3 | Mech. Rotational |
| 4 | Gyrator | 4 | Hydraulic/Fluid |
| 5 | D-Type Element | 5 | Thermal |
| 6 | T-Type Element | | |
| 7 | T-Source | | |

The state-space model output by the MATLAB toolbox is as follows:

$$\begin{bmatrix} \dot{v}_m \\ \dot{F}_K \end{bmatrix} = \begin{bmatrix} -(R \cdot A_p^2 + b)/m & -1/m \\ K & 0 \end{bmatrix} \begin{bmatrix} v_m \\ F_K \end{bmatrix} + \begin{bmatrix} -A_p/m \\ 0 \end{bmatrix} P_s(t)$$

$$v_m = \begin{bmatrix} 1 & 0 \end{bmatrix} \begin{bmatrix} v_m \\ F_K \end{bmatrix} + [0]P_s(t)$$

The dynamic response of this system for the specified output (velocity of the mass element) is shown in Fig. 2. This system was evaluated using a 100 kPa pressure source, a pipe resistance of 100 Pa·s/m³, a sliding resistance of 50 N·s/m, and a spring coefficient of 150 N/m for a mass of 100 kg and a piston with a 10 cm diameter.

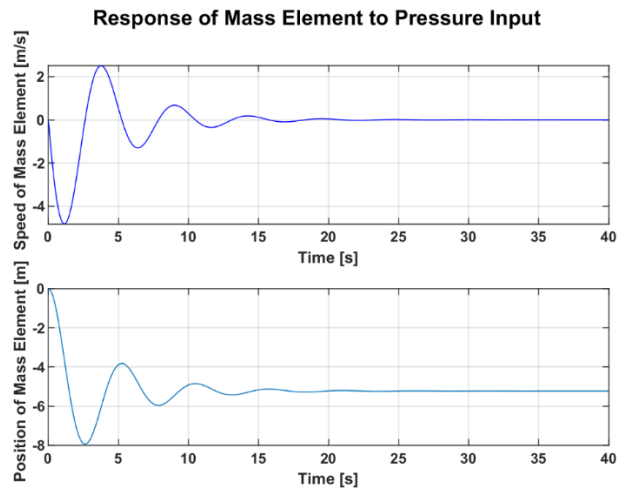

**Fig. 2** Dynamic response of the hydraulically actuated mass system

By observing the velocity response of the mass element, it can be determined that the force of the piston, created via the pressure source, accelerates the block towards the spring element in the negative direction (shown as negative velocity). Once the compression of the spring becomes too great, the block is then pushed back towards the piston briefly until the velocity oscillations settle. By observing the position response of the mass element, which was found by integrating the velocity response, it can be seen that the piston pushes the block approximately 8 meters before the piston force is eventually equalized with the spring force and the mass settles at around the 5-meter position.

As can be observed by the results obtained in this example, the LGtheory MATLAB Toolbox is a simple and robust method of evaluating multi-domain dynamic systems using the LG approach. This toolbox will be utilized to facilitate the design evolution of electronic filter circuits by evaluating the dynamic frequency response of the generated system models.

## 2.2. Genetic Programming

Genetic Programming (GP) is a machine learning technique which generates computer programs consisting of various functions and terminals in an evolutionary manner based on the concept of natural evolution. In natural evolution, members of a population adapt to changing environments over numerous generations, with some members who inherit or adapt beneficial characteristics being more likely to survive, reproduce and, therefore, pass on these benefits, while other members who do not are more likely to die off before reproduction. This process of natural selection means that, over time, the weaker members of the population are unable to pass on their unbeneficial characteristics, thus strengthening the total population.

GP represents members of the population in the form of tree structures, consisting primarily of two main components: functions, which are sub-programs that conduct a specific procedure or routine based on input values and return or perform some output action; and terminals, which can either be functions, having no inputs, or operands, which provide input values to other functions in the tree. The topological structure of these trees is represented by nodes which denote the aforementioned functions and terminals, and lines, which represent the hierarchical connections between nodes. In the tree structure, all inner nodes must be functions as they can accept input actions from other nodes and provide an output action, whereas all outermost nodes of the tree must be terminals as they can only provide output actions.

All functions and terminals used in GP must satisfy the property of closure, which requires that all functions take into consideration type consistency and evaluation safety (Poli et al. 2008). This is important because as the GP process adapts and manipulates trees stochastically through crossover and mutation operations, it is possible that any combination of functions and terminals may occur; thus, consistency in function input/output types, or a method of predictably converting types, is required to ensure that any possible tree structure can be evaluated properly. Similarly, evaluation safety is required as some functions may fail during runtime under certain conditions (i.e., a divide function may fail if the operand of the denominator is 0). It is common practice to ensure that these functions are protected by checking for conditions that would lead to failure and, if so, perform some predetermined output action instead (e.g., if dividing by 0, always return 1). Alternatively, it is possible to allow these failures to occur while applying a large fitness penalty to these solutions in order to potentially eliminate them from the population in subsequent generations.

### 2.2.1. Fitness Function

Similar to how the environment tests an individual's ability to survive in natural evolution, the fitness function in GP evaluates a solution's ability to produce a desired result. This method of evaluation can be conducted in any number of ways that the user sees fit, and can optimize for either maximizing or minimizing the resulting fitness values. One of the most common ways to determine fitness is to calculate the absolute error between the produced result and the desired outcome. For this particular type of fitness measurement, it is desired to minimize the fitness of each member as the fitness value relates directly to the total error for that individual. By prioritizing the minimization of fitness for this type of fitness function, preceding generations will strive to achieve increasingly lower fitness values, thus, minimizing the error. Ultimately, after every member of each generation has been produced and evaluated, the individual with the most desirable fitness value, be it lowest or highest, is chosen as the solution to the GP problem.

### 2.2.2. Selection

Once the fitness of the population has been evaluated, the selection process occurs. Again, this process can be conducted in a number of ways depending on the specific application, but one of the most common methods used is the roulette wheel approach. This method involves spinning a roulette wheel where each member of the population occupies a portion of the wheel with a size corresponding to their fitness with respect to the total fitness of the population. This means that members with more desirable fitness values will be more likely to be selected for reproduction than members with less desirable fitness values. Other methods of selection include variations of the roulette wheel approach, and the tournament selection approach, which involves pitting individuals against each other randomly, with the individual selected to reproduce being the one with the most desirable fitness.

### 2.2.3. Genetic Operations

In GP, two primary genetic operations (and many variants of these operations) are utilized during the reproduction process in order to introduce new individuals and diverse characteristics into the population. These operations are crossover and mutation:

The *crossover* operation occurs between two individuals that have been selected as parents for reproduction. A node is selected from each of these parents and the associated branches are swapped with one another. This process results in two new children who differ slightly from their parents in order to further explore the solution space.

The *mutation* operation involves one individual who has been selected to be a parent. A random node is selected in this individual and the associated tree branch is replaced with a random branch that is created using the available functions and terminals. This process results in one new child who differs from its parent, and is beneficial for introducing more solution diversity into the population.

### 2.2.4. GPLAB

GPLAB is a GP-based MATLAB toolbox developed by Sara Silva at the University of Coimbra, Portugal, which was released for public use in July of 2003 [17]. This toolbox has been under consistent development since its initial release, with the most recent version at the time of writing, version 4.04, being released in June of 2018. While GPLAB's default configuration is for more basic tasks such as symbolic regression, it is also possible to customize various aspects of the toolbox, such as functions and terminal sets, fitness functions, selection procedures, genetic operations, and so on, in order to facilitate the needs of more advanced research.

Since GPLAB is written in MATLAB, along with other benefits such as its robustness and flexibility for user modification, it can be easily adapted to be compatible with any other MATLAB-based scripts or toolboxes such as our LGtheory MATLAB Toolbox. Due to these reasons, GPLAB was selected for implementation with LGtheory for the present work.

## 3. Design Evolution of Electronic Filters

Electronic filters are signal processing electric circuits consisting of discrete electrical components with the purpose of attenuating unwanted frequencies while maintaining the wanted ones. There are many types of electronic filters with different functions and benefits. The purpose of the present section is to explore the application of LG and GP in the design of linear passive low-pass, high-pass, and band-pass filter circuits, and thereby validate the proposed method. Passive filters are filter circuits that contain passive electrical components, such as capacitors, inductors, and resistors. They do not contain active components, such as operational amplifiers, which require external power sources to operate.

### 3.1. Embryo Model

The design evolution process begins with an initial embryo model in order to provide a basis on which the evolutionary process can commence. For all three filters that will be designed in the following sections, the embryo model will consist of a voltage source, $V_S$, in series with a resistive element, $R_S$, representing the source internal resistance, and another resistive element, $R_L$, representing the load resistance, which will be placed in parallel with the last element added to the model between the $n^{th}$ node and the ground node, 1. The values pertaining to the source voltage and the resistances of the embryo model are specified by the user during setup in order to accommodate the specific needs of the designed system. Fig. 3 represents the embryo system model; the evolved filter circuit is constructed within this embryo between the nodes highlighted with the red squares, 3 and *n*. In order to evaluate the evolved models, the voltage of the load resistance element is selected as the output of the state-space model and the frequency response of this element is observed via a Bode plot for each iteration of the evolutionary design process.

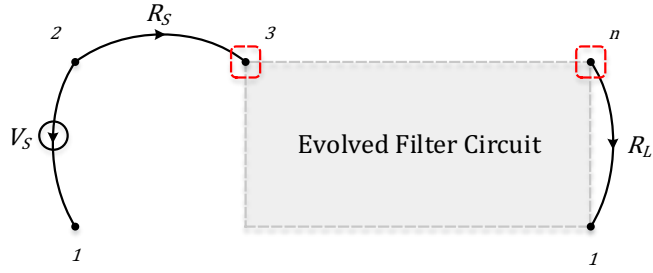

**Fig. 3** Embryo model for evolved filter circuits

## 3.2. Functions

The following sections describe the functions utilized by the GP toolbox to facilitate the construction of the LG models of the filter circuits.

### 3.2.1. Series

The series function takes two input actions and constructs the resulting sub-models in series. In the simplest case, if both actions are the addition of elements to the model, the series function will place both of these elements in series with one another in the system model. For more complex cases, this function will produce the sub-models that result from the associated input branches in such a way that the resulting system components are constructed in series with one another as well.

### 3.2.2. Split

The split function takes two input actions and constructs the resulting sub-models independently of one another. After a split occurs, each resulting sub-model will be constructed separately from one another and grounded independently, resulting in each sub-model being parallel with one another between the node where the split occurs and the ground node. The splitting function is important for the design of electronic filters as it results in a ladder-like topology that is required for constructing higher order filter circuits.

### 3.3. Terminals

The Add_A, Add_D, and Add_T terminals each add their respective passive elements to the embryo system model. These terminals perform the addition of their respective elements in an identical manner, with the exception of the element type index value which must correspond with type of element that is being added to the system. When an element is added, a new node is also created, upon which the model can be built further. If this newly created node is not to be further built upon during the GP process, it will be converted into a ground node. Additionally, these terminals randomly select the parameter values associated with the newly added element within a range of values which are determined based on the source/load resistance values and the cutoff frequencies specified by the user. This parameter selection process is done in order to ensure that the GP fitness will converge toward an optimal value quickly by reducing the solution space which the algorithm must explore.

### 3.4. Fitness Function

Each model generated by the GP algorithm is subsequently sent to the LGtheory toolbox in order to extract the state-space model of the system. If this toolbox determines that it is impossible to extract the state-space models of the evolved systems due to any issues with the model construction by the GP (e.g., excess or lack of state variables, incomplete models, etc.) an error message is returned to the fitness function by the toolbox. Upon detection of such error messages, the fitness function will end for that particular system and a large penalty value shall be applied to the fitness of that individual.

If a state-space model can be extracted, the frequency response of the system is evaluated using the "bode" function in MATLAB. From the data produced by the "bode" function, the magnitude of the output voltage at each point along the plot is evaluated against the desired voltage output for that specific frequency. Ideally, this means that for frequencies that are desired to be passed, the magnitude of the output voltage will be equal to the input voltage, and for unwanted frequencies, the magnitude of the output voltage will be equal to zero. The absolute error values found between the desired and the actual voltage values within the entire frequency range are summed and equated to the fitness of the evolved system. In band pass filter evolution, the absolute error values associated with the pass band frequencies are squared to increase their impact on the fitness value. This is due to the pass band generally being narrow compared to the total frequency range; squaring the error of the pass band helps to equalize the algorithm's prioritization of the pass band and attenuated frequencies, thus tending to generate a more optimal system.

The GP algorithm is run for the specified amount of generations and population size, and the evolved model that results in the lowest fitness value is selected and presented to the user as the final design.

### 3.5. Interpreting Tree Structures

Fig. 4 shows the tree structure evolved for the low pass filter model. The GP algorithm executes functions in a manner that the first sub-tree from the function node is evaluated in its entirety before the second sub-tree is considered. This is represented in the figure by the arrows lining the perimeter of the tree in a counter-clockwise direction, which represents the order in which the terminal nodes are executed, and thus, the order in which elements are added to the system.

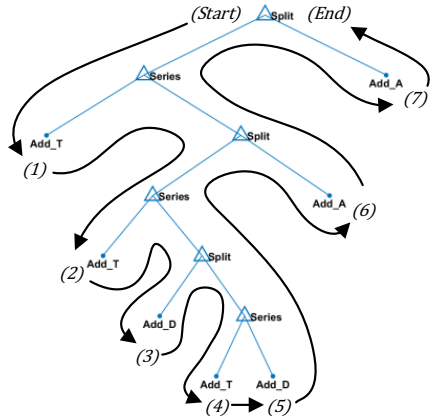

**Fig. 4** Tree structure of the evolved low pass filter with the directionality of node execution

This tree structure is interpreted as follows: at (Start), a split is created at node 3 of the embryo model. At (1), an inductor (T-Type) element is added to the model at node 3 and a new node, 4, is created. Another split is created at node 4, and at (2), a second inductor element and node, 5, are added to the model. At this newly created node, another split occurs with the addition of a resistive (D-Type) element at (3), with the node created by the addition of this element being connected to ground. A series connection of the third inductor and the second resistor elements are added to node 5 at (4) and (5), creating new nodes 6 and 7. Following the arrow from (5) to (6) results in the grounding of node 7, and the addition of a capacitor (A-Type) element between the previously split node 4 and a new node, which is subsequently grounded. Finally, following the arrow from (6) through (7) to (End) results in the addition of a second capacitor element between node 3 and ground.

After the tree has been interpreted by the GP algorithm and the LG model constructed, the load resistance element is added to the system between the highest integer node, in this case 6, and ground. This results in the final low pass filter LG model, as shown in Fig. 6.

## 4. Results and Discussion

### 4.1. Low Pass Filter Evolution

The low pass filter evolution program was run for a system containing a 10-volt source with an internal resistance of 750 Ohms, and a load resistance of 50 Ohms. The filter was designed for a cutoff frequency of 50 kHz over 100 generations with a population size of 50. Fig. 4 of the previous section shows the tree structure that was constructed by the GP algorithm for these input values.

The two graphs shown in Fig. 5 demonstrate the evolution of the fitness and structural complexity, respectively, of the evolved tree structures over the course of the 100 generations. The left graph shows the change in the fitness of the "best so far" tree structure, as well as, the median and average finesses, and the average standard deviations of all trees for each generation of the evolution process. Similarly, the right graph shows the evolving structural complexity of the "best so far" solution in the form of depth, size, and percentage of introns of the associated tree structure.

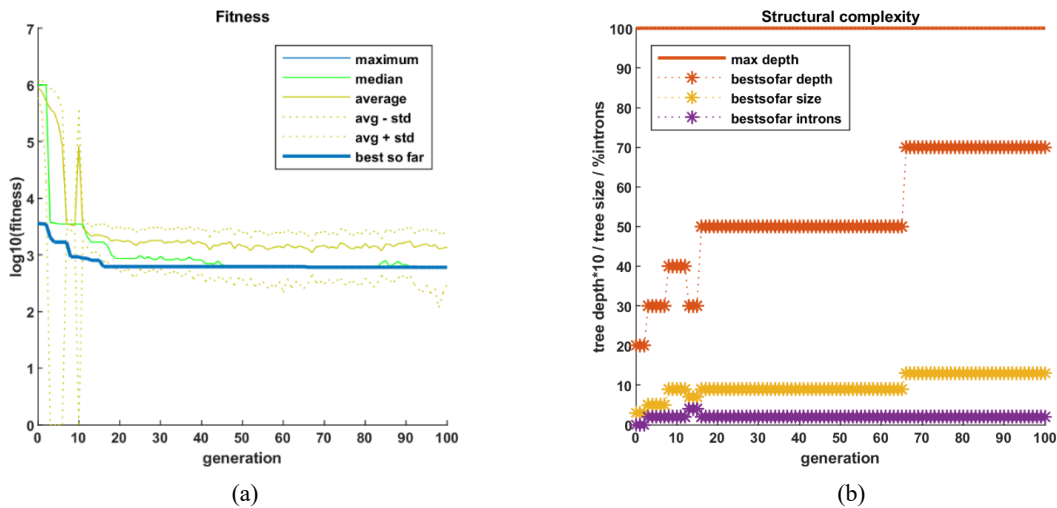

**Fig. 5** (a) Fitness and (b) structural complexity for the evolution of the low pass filter

The resulting LG model and the equivalent circuit diagrams are shown in Fig. 6.

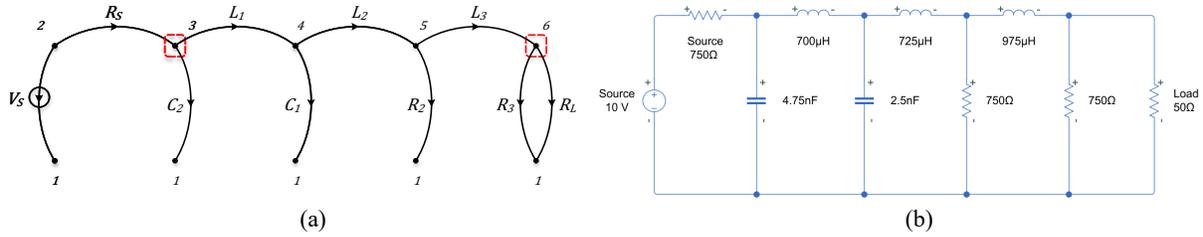

(a)            (b)

**Fig. 6** (a) LG model and (b) circuit diagram of the evolved low pass filter

The frequency response of the system is shown in Fig. 7.

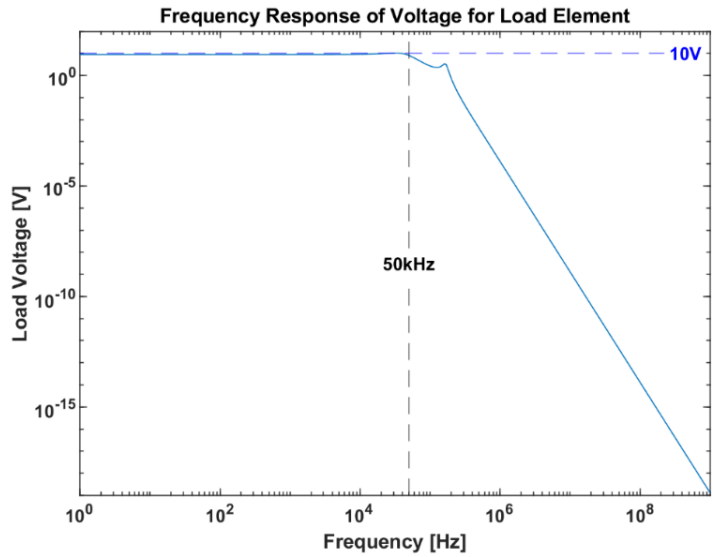

**Fig. 7** Frequency response function of the evolved low pass filter

## 4.2. High Pass Filter Evolution

The high pass filter evolution program was run for a system containing a 10-volt source with an internal resistance of 750 Ohms, and a load resistance of 50 Ohms. The filter was designed for a cutoff frequency of 300 kHz over 100 generations, with a population size of 50. Fig. 8 shows the tree structure that was constructed by the GP algorithm for these input values.

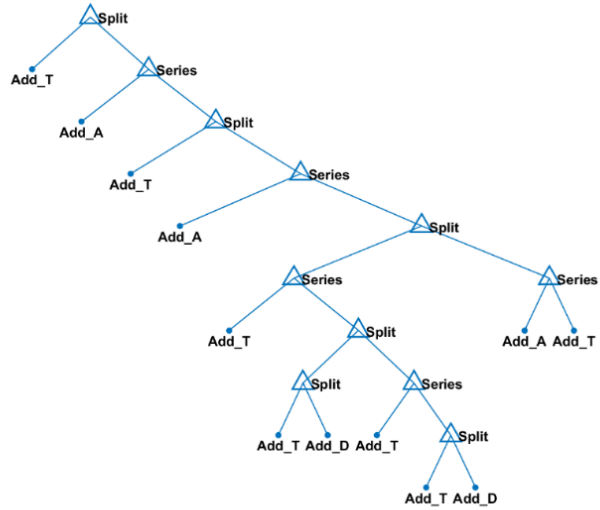

**Fig. 8** Tree structure of the evolved high pass filter

The two graphs shown in Fig. 9 demonstrate the evolution of the fitness and structural complexity of the evolved tree structures over the course of the 100 generations.

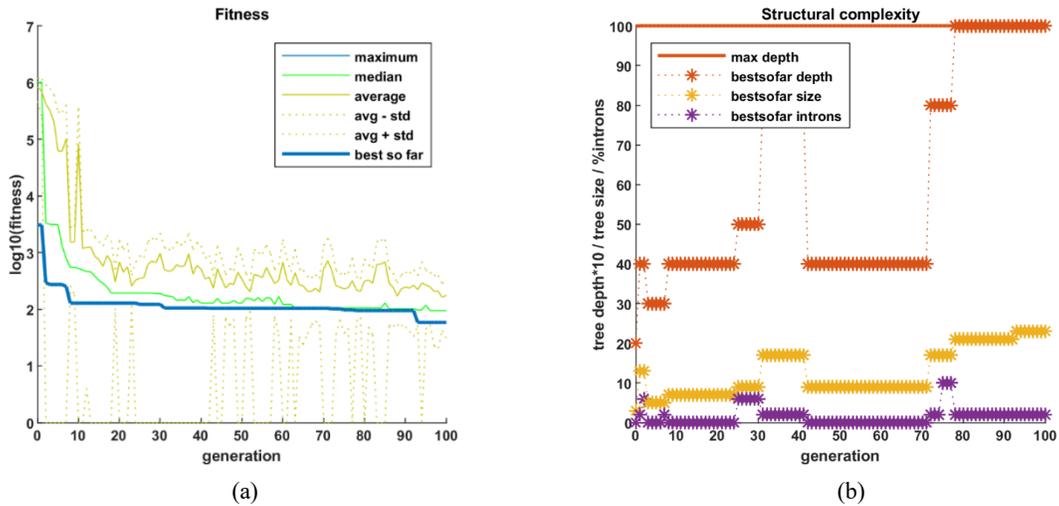

**Fig. 9** (a) Fitness and (b) structural complexity for the evolution of the high pass filter

The resulting LG model and the equivalent circuit diagrams are shown in Fig. 10.

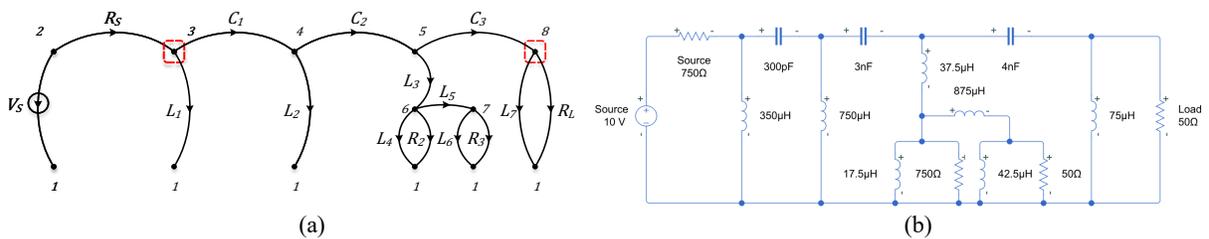

**Fig. 10** (a) LG model and (b) circuit diagram of the evolved high pass filter

The frequency response function of the system is shown in Fig. 11.

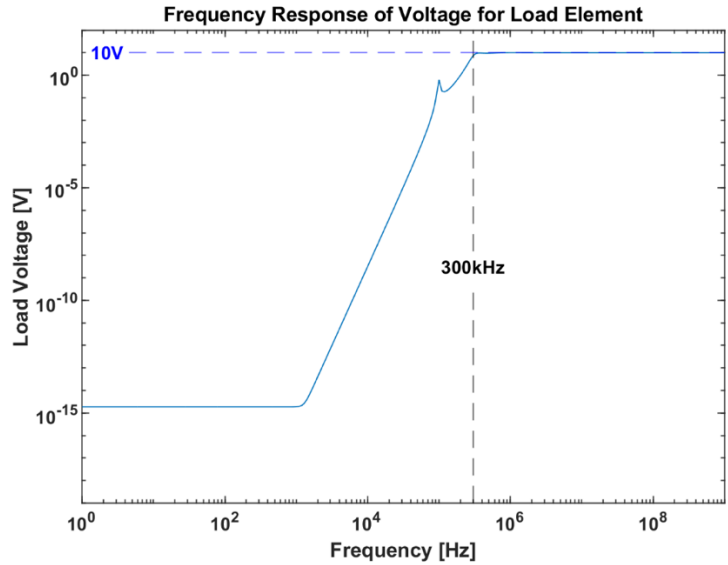

**Fig. 11** Frequency response function of the evolved high pass filter

## 4.3. Band Pass Filter Evolution

The band pass filter evolution program was run for a system containing a 10-volt source with an internal resistance of 750 Ohms, and a load resistance of 50 Ohms. The filter was designed for a lower cutoff frequency of 20 kHz and an upper cutoff frequency of 250 kHz, over 100 generations with a population size of 50. Fig. 12 shows the tree structure, which was constructed by the GP algorithm for these input values.

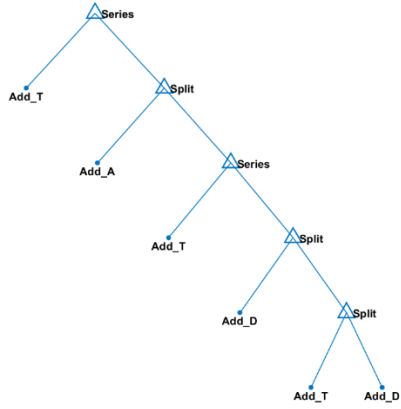

**Fig. 12** Tree structure of the evolved band pass filter

The two graphs shown in Fig. 13 demonstrate the evolution of the fitness and structural complexity of the evolved tree structures over the course of the 100 generations.

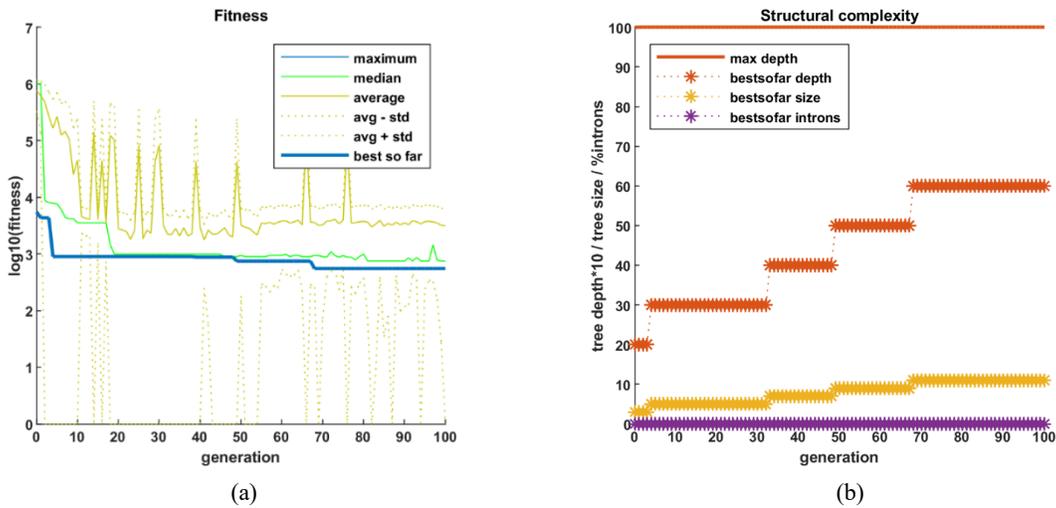

**Fig. 13** (a) Fitness and (b) structural complexity for the evolution of the band pass filter

The resulting LG model and equivalent circuit diagrams are shown in Fig. 14.

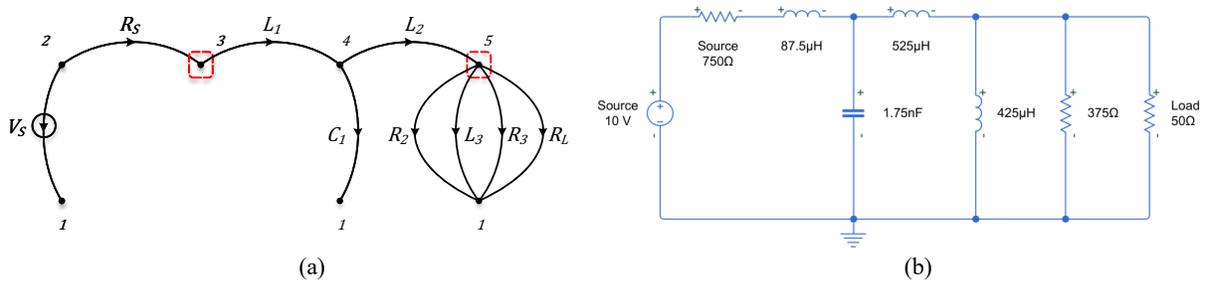

**Fig. 14** (a) LG model and (b) circuit diagram of the evolved band pass filter

The frequency response function of the system is shown in Fig. 15.

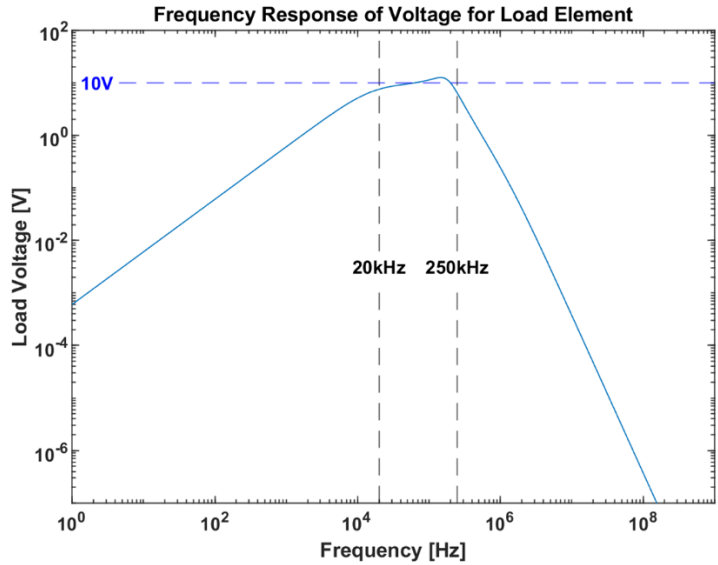

**Fig. 15** Frequency response function of the evolved band pass filter

## 5. Conclusion

This paper presented the application of the combined LG modeling and GP design approach for the evolutionary design of passive electronic filter circuits. While the use of GP for the design of electronic filters is not new, this application of design evolution was utilized in order to demonstrate the capabilities of LG modeling and the LGtheory MATLAB Toolbox for automated design by evaluating the evolved systems via dynamic simulation.

In this paper, three types of passive filter circuits (low pass, high pass, and band pass filters) were designed using the combination of LG modeling and GP. The results of the frequency response of these systems demonstrated that the LG/GP program was able to successfully evolve high-order filter circuits that were effective at attenuating the voltages of undesirable frequencies while maintaining the voltages of desired frequencies.

While the present application of designing filter circuits is only concerned with the electrical energy domain, due to the unified and integrated nature of LG modeling, a characteristic which other single-domain evaluation tools lack, it is possible to extend the present work to other domains, and even combinations of different domains, for the design of complex multi-domain mechatronic system. The present paper further validates the capabilities of the LGtheory MATLAB Toolbox, which has been developed by us, as a robust and reliable software package for the design and simulation of complex dynamic systems.

## ACKNOWLEDGMENTS

This research was funded by the NSERC Discovery Program, grant number RGPIN-2017-05762.